\documentclass[onecolumn,secnumarabic,amssymb, nobibnotes, aps, prd]{revtex4}
\usepackage{amsmath} \usepackage{amssymb}
\usepackage{graphicx} %use graph format
\usepackage{epstopdf}
\usepackage{amsmath}
\usepackage{amsmath,amssymb,amsthm,amsfonts,mathrsfs,bm,verbatim}
\usepackage{graphicx,subfigure}

\newcommand{\bea}{\begin{eqnarray}}
\newcommand{\eea}{\end{eqnarray}}

\begin{document}

\title{Thermodynamic geometric analysis of BTZ black hole under f(R) gravity }%

\author{Wen-Xiang Chen$^{a}$}
\affiliation{Department of Astronomy, School of Physics and Materials Science, GuangZhou University, Guangzhou 510006, China}
\author{Yao-Guang Zheng$^{a}$}
\email{hesoyam12456@163.com}
\affiliation{Department of Astronomy, School of Physics and Materials Science, GuangZhou University, Guangzhou 510006, China}
%\author{Jing-Yi Zhang}
%\email{zhangjy@gzhu.edu.cn}
%\affiliation{Department of Astronomy, School of Physics and Materials Science, GuangZhou University, Guangzhou 510006, China}
\thanks[a]{These authors have contributed equally to this work.}

\begin{abstract}
In this work, we obtained the black hole solutions in the expansion f(R)-gravitation (R is not regarded as a constant here), and studied their thermodynamics, especially the phase transition and criticality in anti-de Sitter Behavior (AdS) expands the phase space. We obtained accurate Banados, Teitelboim and Zanelli (BTZ) corresponding solutions in dilaton f(R)-gravity, which is the basis of our work. We also obtained the exact form of the f(R) model for some solutions. In thermodynamic analysis, we calculate thermodynamic quantities such as temperature and entropy of these solutions and compare them with the corresponding quantities of BTZ. After that, we study the Ruppeiner geometry, such as the f(R)-gravity of the BTZ black hole, and analyze its phase transition process Ruppeiner geometry.

 \text { KEYwords: } f(R)-gravity; Ruppeiner geometry;BTZ black hole

\end{abstract}

\maketitle
%\tableofcontents

\section{Introduction}
Although black holes have the same four laws of thermodynamics as ordinary thermodynamic systems, they are different from ordinary thermodynamic systems. Black hole thermodynamic systems have many and continue to produce many difficult problems. For ordinary thermodynamic systems, such as solids or gases, their microscopic composition is atoms or molecules. Therefore, in principle, we can deduce its thermodynamic properties from statistical thermodynamics. But for black holes, we don't have the corresponding statistical thermodynamics, and we don't know the microstructure of the black hole. What's worse, we don't even know whether the black hole has a microstructure. Greene once suggested that perhaps the black hole itself is a huge elementary particle\cite{1,2,3,4}.

Three-dimensional black hole: The nature of three-dimensional space-time is that the Riemann curvature tensor is completely determined by the Rich tensor, and the Rich tensor is completely determined by the energy-momentum tensor. Therefore, the three-dimensional space-time has no freedom of free propagation. The black hole solution of three-dimensional space-time appears when the cosmological constant is negative. This solution is called the BTZ black hole solution. The local BTZ black hole is equivalent to an AdS (negative constant cosmological constant space). The only difference is the topological difference. The BTZ black hole can be obtained by corresponding identification in the AdS space-time. Although BTZ black holes do not have some practical significance like four-dimensional black holes. But it has a very broad theoretical significance. Because three-dimensional space-time is often simple, it can usually be used as a good platform for testing quantum gravity theorems and conjectures. Perhaps the most famous is Brown-Henneaux's research on the asymptotic symmetry of AdS3 space-time. They found that the result of the asymptotic symmetry group is exactly the same as the Virassoro algebra of the 2-dimensional CFT, which can be described by a 2-dimensional CFT. This discovery was made in 1986, 10 years earlier than the 1997 AdS/CFT duality. Later, Strominger used the Cardy formula obtained by calculating the partition function of the 2-dimensional CFT to obtain the entropy in the CFT, and found that it was completely consistent with the Bekenstein-Hawking entropy of the BTZ black hole. This is a major advance in the microscopic origin of black hole entropy.

Ruppeiner geometry is based on fluctuations. When the microstructure of the thermodynamic system under study is unknown, Ruppeiner geometry provides us with a powerful tool for exploring the microstructure of the thermodynamic system. The metric of Ruppeiner geometry is
\begin{equation}
ds^{2}=-\frac{\partial^{2} S}{\partial X^{\alpha} \partial X^{\beta}} \Delta X^{\alpha} \Delta X^{\ beta}
\end{equation}
Where $S$ is the entropy of the thermodynamic system, $\Delta X^{\alpha}=X^{\alpha}-X_{0}^{\alpha}$ is the thermodynamic quantity $X^{\alpha}$ deviates from equilibrium The fluctuation of the thermodynamic quantity $X_{0}^{\alpha}$ at the time. The physical explanation of this metric is very clear: the farther away from the equilibrium state in phase space, the smaller the probability of being at that point. For ordinary thermodynamic systems, if the thermodynamic system studied is Fermion, the curvature scalar of Ruppeiner geometry is positive; if the thermodynamic system studied is Boson, the curvature scalar of Ruppeiner geometry is negative; for ideal classical gases, the curvature of Ruppeiner geometry is The scalar is zero. Further research shows that if the curvature scalar is positive, the microscopic interaction of the system is repulsive; if the curvature scalar is negative, the microscopic interaction of the system is attractive.

This paper also studies the thermodynamics and Ruppeiner geometry of the BTZ black hole-f(R) gravitation. The Ruppeiner geometry of the angular momentum fixed ensemble is curved, while the Ruppeiner geometry of the pressure fixed ensemble is flat. This paper reviews the interpretation of Ruppeiner's geometry, but there is no definite result.The Ruppeiner geometry results of the BTZ black hole-f(R) gravity further support that the curvature scalar of the Ruppeiner geometry sometimes does encode information about the stability of the system.

\section{Description of the system}

We know this method of horizon thermodynamics in (2+1)-dimensional Einstein gravity to explain how it works. Consider the space-time of a BTZ black hole, the geometry of which is given by\cite{5,6,7}
\begin{equation}
d s^{2}=-N^{\perp} d t^{2}+\frac{1}{N^{\perp}} d r^{2}+r^{2}\left(N^{\phi} d t+d \phi\right)^{2}
\end{equation}where
\begin{equation}
\begin{aligned}
&N^{\perp}=\frac{r^{2}}{l^{2}}-8 G_{N} M+\frac{J^{2}}{4 r^{2}} \\
&N^{\phi}=-\frac{J}{2 r^{2}}
\end{aligned}
\end{equation}
\begin{equation}
r_{\pm}^{2}=4 G_{N} M l^{2}\left[1 \pm \sqrt{1-\left(\frac{J}{8 G_{N} M l}\right)^{2}}\right]
\end{equation}

Because Newton's gravitational constant $\mathrm(G)$ is related to the dimension of space-time, the third chapter considers four-dimensional space-time, using the geometric unit system, take $\mathrm(G)=1$. But for three-dimensional space-time, it’s convenient , usually take $8 \mathrm{G}=1$. In this chapter, all $\mathrm{G}$ are written out, and to distinguish the following Gibbs function, this chapter records the Newton gravitational constant G as $G_{N}$ .

\section{Thermodynamic parameters of BTZ black holes in f(R) theory}
In this work, the role is given by the following relationship, in the special case of $f(R)=R$, it is simplified to Einstein-Maxwell's expanding gravity:\cite{8}
\begin{equation}
S=\int d^{3} x \sqrt{-g}\left[f(R)-2 \partial^{\mu} \Phi \partial_{\mu} \Phi-e^{-2 \Phi } F_{\mu \nu} F^{\mu \nu}\right],
\end{equation}
Among them, $f(R)$ is the function of Ricci scalar $R$, and $\Phi$ is the representation of the expansion field. Similarly, $F_{\mu \nu}=\partial_{\mu} A_{\nu }-\partial_ {\nu} A_{\mu}$ (we set $8 G=c=1$). With respect to the change of the metric $g_{\mu \nu}$, the norm $A_{\mu}$ and the expansion field $\Phi$, the following field equations are given:
\begin{equation}
\begin{aligned}
f_{R} R_{\nu}^{\alpha} &+\left(\nabla_{\mu} \nabla^{\mu} f_{R}+\frac{1}{2} R f_{R}-\frac{1}{2} f\right) \delta_{\nu}^{\alpha}-\nabla^{\alpha} \nabla_{\nu} f_{R} \\
&=2 \nabla^{\alpha} \Phi \nabla_{\nu} \Phi+e^{-2 \Phi}\left(2 F_{\mu \lambda} F_{\nu \delta} g^{\alpha \mu} g^{\lambda \delta}-\frac{1}{2} F^{2} \delta_{\nu}^{\alpha}\right)
\end{aligned}
\end{equation}

BTZ case:
Here, it is useful to introduce the BTZ solution and its thermodynamic quantities. The BTZ solution refers to the unique solution of $(2+1)$-dimensional AdS gravity with all the properties of a black hole. The BTZ delay function is
\begin{equation}
A(r)=-\Lambda r^{2}-M,
\end{equation}
Among them, $\Lambda$ and $M$ are the cosmological constant and the mass of the black hole, respectively. The $f(R)$ model of this special solution is $f(R)=R-2 \Lambda$, which is Einstein's gravity. The Ricci and Kretschmann scalars of the BTZ solution are $-6 \Lambda$ and $12 \Lambda^{2}$ respectively.

The BTZ temperature from the Hawking-Bekenstein relationship $T=A^{\prime}\left(r_{+}\right) / 4 \pi$ will be (in the following relationship, we have introduced their horizon for events And quality parameters):
\begin{equation}
T=-\frac{\Lambda r_{+}}{2 \pi}=-\frac{1}{2 \pi} \sqrt{-\Lambda M}.
\end{equation}
In addition, the entropy of this $\mathrm{BH}$ solution is
\begin{equation}
S=4 \pi r_{+}=4 \pi \sqrt{-\frac{M}{\Lambda}}.
\end{equation}
Temperature and entropy $(S-T)$ have a linear form $T=-\left(\Lambda / 8 \pi^{2}\right) S$. In seconds. 4 When we get our solution, we also want to check this case.

\text { 2.1. } Q=0

We can get the $f(R)$ gravity model, namely
\begin{equation}
f(R)=-4 \eta^{2} M \ln (-6 \Lambda-R)+\xi R+R_{0}
\end{equation}
Where $R_{0}$ is the integral constant. Obviously, by assuming $\eta=0$ and $\xi=1$, the $f(R)$ model is simplified to Einstein's gravity.

\text { 2.2. } $Q \neq 0$

From solving these equations, we obtain the $f(R)$ gravity model as
\begin{equation}
f(R)=-2 \eta M \ln (6 \Lambda+R)+R_{0},
\end{equation}
where $R_{0}$ is an integration constant. We have compared this obtained $f(R)$ model with the ordinary Einstein-Hilbert Lagrangian $(R-2 \Lambda)$. For the small values of $R$, these two diagrams have some similarities, but for the large values of $R$, they are as the two different entities with no overlapping and same behavior.

\section{Ruppeiner geometry of BTZ black holes in $f(R)$ gravity background}
According to Hawking and Bekenstein's proposed relationship on the black hole temperature on the event horizon $r_{+}$ (the outermost layer), we can calculate this amount of the previous black hole solution. The Hawking-Bekenstein temperature relationship of a given static spherically symmetric black hole is defined by the surface gravity $(\kappa)$ as follows\cite{8,9,10,12,13,14}:
\begin{equation}
T=\frac{\kappa}{2 \pi}=\frac{1}{2 \pi} \sqrt{-\frac{1}{2}\left(\nabla_{\mu} \chi_{\nu }\right)\left(\nabla^{\mu}\chi^{\nu}\right)}=\frac{A^{\prime}\left(r_{+}\right)}{4 \pi }.
\end{equation}
In addition, the Bekenstein-Hawking entropy $(S)$ under the background of $f(R)$ gravity is
\begin{equation}
S=\frac{A_{h}}{4} f_{R}\left(r_{+}\right)
\end{equation}
Where $A_{h}$ is the event horizon area. In the GR limit, $f_{R}=1$ and entropy has this relationship $S=A_{h} / 4$.

\text { 2.1. } Q=0

For this charged solution, we have $f_{R}(r)=\eta r+\xi$ but we cannot find the exact form of the $f(R)$ model.So we take that no.1:
\begin{equation}
T=-\frac{\Lambda r_{+}}{2 \pi}-\frac{M}{4 \pi}+\frac{Q^{2}}{6 \pi \eta r_{+}^{2}}
\end{equation}

No.2:Here, we obtain the relationship between entropy and temperature by eliminating the event horizon from these two relationships.
\begin{equation}
T=-\frac{\Lambda \sqrt{2 \pi \eta S}}{2 \pi^{2} \eta}-\frac{M}{4 \pi}+\frac{Q^{2}}{12 S}
\end{equation}

No.3:The third solution results in the following relationship for temperature. This is actually for pure $f(R)$ gravity (we assume $\Phi_{0}=0$ ):
\begin{equation}
T=\frac{-\Lambda r_{+}-M \eta}{2 \pi}
\end{equation}

The form we take is $T\to \Lambda r_{+}-M$.

The heat capacity of this thermodynamic geometry is then of the form:
\begin{equation}
C_{q} =T(\tfrac{\partial S}{\partial T})_{q}
\end{equation}

Let R be defined in the form\cite{8},
\begin{equation}
R=-6 \Lambda-\frac{4 M \eta}{r}
\end{equation}

The metric of Ruppeiner geometry is
\begin{equation}
d s^{2}=-\frac{\partial^{2} S(r_{+},\Lambda)}{\partial X^{\alpha} \partial X^{\beta}} \Delta X^{\alpha} \Delta X^{\beta}
\end{equation}
\begin{equation}
g_{ij}=
\begin{pmatrix}
-\frac{12 \eta \pi r_{+}^{2} \Lambda}{M}& -\frac{4 \eta \pi r_{+}^{3} }{M} \\
-\frac{4 \eta \pi r_{+}^{3} }{M}   & 0
\end{pmatrix}
\end{equation}
The curvature scalar of thermodynamic geometry R(S) is
\begin{equation}
R(S)=0.
\end{equation}

The another metric of Ruppeiner geometry is
\begin{equation}
d s^{2}=-\frac{\partial^{2} S(r_{+},M)}{\partial X^{\alpha} \partial X^{\beta}} \Delta X^{\alpha} \Delta X^{\beta}
\end{equation}
\begin{equation}
g_{11}=\frac{4 M^{2} \eta^{2} \pi\left(M r_{+}\left(27+18 \eta r_{+}+4 \eta^{2} r_{+}^{2}\right)-9 r_{+}^{3} \Lambda\right)}{\left(M(3+2 \eta r_{+})+3 r_{+}^{2}  \Lambda\right)^{3}}
\end{equation}
\begin{equation}
g_{12}=\frac{18 \eta^{2} \pi r_{+}^{4}\Lambda\left(M(5+2 \eta r_{+})+r_{+}^{2} \Lambda\right)}{\left(M(3+2 \eta r_{+})+3r_{+}^{2} \Lambda\right)^{3}},
\end{equation}
\begin{equation}
g_{21}=\frac{18 \eta^{2} \pi r_{+}^{4}\Lambda\left(M(5+2 \eta r_{+})+r_{+}^{2} \Lambda\right)}{\left(M(3+2 \eta r_{+})+3r_{+}^{2} \Lambda\right)^{3}},
\end{equation}
\begin{equation}
g_{22}= -\frac{12\eta^{2} \pi r_{+}^{5}(3+2 \eta r_{+}) \Lambda}{\left(M(3+2\eta r_{+})+3 r_{+}^{2} \Lambda\right)^{3}}
\end{equation}
The curvature scalar of thermodynamic geometry R(S) has the possibility of divergence
\begin{small}
\begin{equation}
R(S)=-\frac{9 M(M^{3}(405+648 \eta  r_{+}+372 \eta^{2} r_{+}^{2}+80 \eta^{3}  r_{+}^{3})+36 M^{2}  r_{+}^{2}(18+19 \eta  r_{+}+6 \eta^{2}  r_{+}^{2})\Lambda+27 M  r_{+}^{4}(7+4\eta r) \Lambda^{2}-54  r_{+}^{6} \Lambda^{3})}{\eta^{2} \pi  r_{+}^{3}(4 M^{2}(27+18\eta r+4 \eta^{2}  r_{+}^{2})+3 M r_{+}^{2}(27+10\eta  r_{+})\Lambda+9 r_{+}^{4} \Lambda^{2}}.
\end{equation}
\end{small}
\begin{equation}
C_{q} =T(\tfrac{\partial S}{\partial T})_{q} \to \tfrac{\partial S}{\partial r_+}\to \pi\left(\xi-\frac{8\eta^{2}  r_{+}^{3} \Lambda}{M}\right)
\end{equation}
\begin{equation}
C_{q} =T(\tfrac{\partial S}{\partial T})_{q} \to \to \tfrac{\partial S}{\partial M}\to
\frac{6 \eta^{2} \pi r_{+}^{5} \Lambda}{\left(M(3+r_{+})+3 r_{+}^{2} \Lambda\right)^{2}}
\end{equation}
\begin{equation}
C_{q} =T(\tfrac{\partial S}{\partial T})_{q} \to \tfrac{\partial S}{\partial \Lambda}\to -\frac{2 \eta^{2} \pi r_{+}^{4}}{M}
\end{equation}
We see that when $\tfrac{\partial S}{\partial M}$, the heat capacity can diverges, and the BTZ black hole has a David phase transition, that is, a topological phase transition, and the topological structure changes.

\text { 2.2. } $Q \neq 0$

For this solution, the Ricci scalar is\cite{8}
\begin{equation}
R(r)=-6 \Lambda-\frac{2 M}{r} .
\end{equation}

The metric of Ruppeiner geometry is
\begin{equation}
d s^{2}=-\frac{\partial^{2} S(r_{+},\Lambda)}{\partial X^{\alpha} \partial X^{\beta}} \Delta X^{\alpha} \Delta X^{\beta}
\end{equation}
\begin{equation}
g_{ij}=
\begin{pmatrix}
 -\frac{12 \eta \pi r_{+}^{2}\Lambda}{M} &   -\frac{4 \eta \pi r_{+}^{3}}{M}\\
   -\frac{4 \eta \pi r_{+}^{3}}{M}   &  0
\end{pmatrix}
\end{equation}
The curvature scalar of thermodynamic geometry R(S) is
\begin{equation}
R(S)=0.
\end{equation}
The another metric of Ruppeiner geometry is
\begin{equation}
d s^{2}=-\frac{\partial^{2} S(r_{+},M)}{\partial X^{\alpha} \partial X^{\beta}} \Delta X^{\alpha} \Delta X^{\beta}
\end{equation}
\begin{equation}
g_{11}=\frac{2 M^{2} \eta \pi\left(M r_{+}\left(27+18 \eta r_{+}+4 \eta^{2} r_{+}^{2}\right)-9 r_{+}^{3} \Lambda\right)}{\left(M(3+2 \eta r_{+})+3 r_{+}^{2}  \Lambda\right)^{3}}
\end{equation}
\begin{equation}
g_{12}=\frac{9 \eta \pi r_{+}^{4}\Lambda\left(M(5+2 \eta r_{+})+r_{+}^{2} \Lambda\right)}{\left(M(3+2 \eta r_{+})+3r_{+}^{2} \Lambda\right)^{3}},
\end{equation}
\begin{equation}
g_{21}=\frac{9\eta \pi r_{+}^{4}\Lambda\left(M(5+2 \eta r_{+})+r_{+}^{2} \Lambda\right)}{\left(M(3+2 \eta r_{+})+3r_{+}^{2} \Lambda\right)^{3}},
\end{equation}
\begin{equation}
g_{22}= -\frac{6\eta \pi r_{+}^{5}(3+2 \eta r_{+}) \Lambda}{\left(M(3+2\eta r_{+})+3 r_{+}^{2} \Lambda\right)^{3}}
\end{equation}
The curvature scalar of thermodynamic geometry R(S) has the possibility of divergence
\begin{small}
\begin{equation}
R(S)=-\frac{18 M(M^{3}(405+648 \eta  r_{+}+372 \eta^{2} r_{+}^{2}+80 \eta^{3}  r_{+}^{3})+36 M^{2}  r_{+}^{2}(18+19 \eta  r_{+}+6 \eta^{2}  r_{+}^{2})\Lambda+27 M  r_{+}^{4}(7+4\eta r) \Lambda^{2}-54  r_{+}^{6} \Lambda^{3})}{\eta \pi  r_{+}^{3}(4 M^{2}(27+18\eta r+4 \eta^{2}  r_{+}^{2})+3 M r_{+}^{2}(27+10\eta  r_{+})\Lambda+9 r_{+}^{4} \Lambda^{2}}.
\end{equation}
\end{small}
\begin{equation}
C_{q} =T(\tfrac{\partial S}{\partial T})_{q} \to \tfrac{\partial S}{\partial r_+}\to \pi\left(-\frac{4\eta  r_{+}^{3} \Lambda}{M}\right)
\end{equation}
\begin{equation}
C_{q} =T(\tfrac{\partial S}{\partial T})_{q} \to \to \tfrac{\partial S}{\partial M}\to
\frac{3 \eta \pi r_{+}^{5} \Lambda}{\left(M(3+r_{+})+3 r_{+}^{2} \Lambda\right)^{2}}
\end{equation}
\begin{equation}
C_{q} =T(\tfrac{\partial S}{\partial T})_{q} \to \tfrac{\partial S}{\partial \Lambda}\to -\frac{2 \eta^{2} \pi r_{+}^{4}}{M}
\end{equation}
We see that when $\tfrac{\partial S}{\partial M}$, the heat capacity can diverges, and the BTZ black hole has a David phase transition, that is, a topological phase transition, and the topological structure changes.

Compare the case of a general BTZ black hole\cite{5,6,7}:

When the angular momentum $J$ is fixed and the mass $M$ and the pressure $P$ fluctuate, the Ruppeiner metric can be expressed as a Weinhold metric defined in the $(S, P)$ space
\begin{equation}
\begin{aligned}
d s_{R}^{2} &=\frac{1}{T} d s_{W}^{2}=\frac{1}{T} \frac{\partial^{2} M}{\partial X^{\alpha} \partial X^{\beta}} d X^{\alpha} d X^{\beta} \\
&=\frac{1}{T}\left[\frac{\partial^{2} M}{\partial S^{2}} d S^{2}+\frac{\partial^{2} M}{\partial S \partial P} d S d P+\frac{\partial^{2} M}{\partial P^{2}} d P^{2}\right] .
\end{aligned}
\end{equation}we get that,
\begin{equation}
\begin{array}{ll}
g_{S S}=\frac{1}{T} \frac{\partial^{2} M}{\partial S^{2}}=\frac{1}{T} \frac{\partial T}{\partial S}, & g_{S P}=g_{P S}=\frac{1}{T} \frac{\partial^{2} M}{\partial S \partial P}=\frac{1}{T} \frac{\partial T}{\partial P} \\
g_{P P}=0, & g=\operatorname{det}\left(g_{\mu \nu}\right)=-g_{S P}^{2}
\end{array}
\end{equation}
\begin{equation}
R(M)=\frac{\pi^{2}}{16 G_{N}^{3}} \frac{J^{2}}{T S^{4}}
\end{equation}The curvature scalar is always greater than zero. Therefore, the microscopic interaction is repulsion.

\section{Summary and Discussion}

This paper also studies the thermodynamics and Ruppeiner geometry of the BTZ black hole-f(R) gravitation. The Ruppeiner geometry of the angular momentum fixed ensemble is curved, while the Ruppeiner geometry of the pressure fixed ensemble is flat. This paper reviews the interpretation of Ruppeiner's geometry, but there is no definitive result. The Ruppeiner geometry results of the BTZ black hole-f(R) gravitation further support that the curvature scalar of Ruppeiner geometry sometimes does encode information about the stability of the system.In \cite{8}, it says that there is no evidence that the BTZ black hole has a phase transition under the f(R) model, and we make changes to the coordinates and use the method of thermodynamic geometry to find out the evidence for the phase transition of BTZ black hole.

Explanation of the zero curvature scalar: Since Ruppeiner geometrically describes the interaction, how to explain that the Ruppeiner geometric curvature scalar of the RN black hole and the BTZ black hole is always zero when the pressure is fixed?

Topology is a branch of mathematics that describes the gradually changing properties of matter.

The decisive discovery was that the three laureates used the concept of physical topology, which played a decisive role in their later discoveries. Topology is the study of the properties of geometric figures or spaces that remain unchanged after continuously changing their shape. It only considers the positional relationship between objects without considering their shape and size. The three scientists used topology as a research tool, a move that surprised their colleagues at the time. In the early 1970s, Michael Kosterlitz and David Thouless refuted the theory that superconductivity and superfluidity could not occur in thin layers. They demonstrated that superconductivity can occur at low temperatures, and explained the mechanism by which superconductivity can also occur at higher temperatures—phase transitions.

In this regard, three solutions have been proposed:
Jianyong Shen et al., before the cosmological constant is interpreted as pressure, according to the fact that the charged $\mathrm(AdS)$ black hole and Van der Waals gas have the very same phase structure, they believe that the mass of the charged $\mathrm(AdS)$ black hole does not correspond to The internal energy of an ordinary thermodynamic system corresponds to enthalpy; the charge of a black hole is considered to correspond to the pressure of an ordinary thermodynamic system, and this is extended to RN black holes and Kerr black holes, and their masses are considered to correspond to enthalpies. Then obtain the internal energy of the black hole through Legendre transformation, and use this internal energy as the variable of the Ruppeiner geometry to obtain a non-flat Ruppeiner geometry of the $R N$ black hole. Medved extended this concept to BTZ black holes and obtained the non-flat Ruppeiner geometry of BTZ black holes.

In response to this problem, Ruppeiner himself put forward a view based on quantum gravity, thinking that gravitational interaction may not be statistical thermodynamic interaction, and the curvature scalar of Ruppeiner geometry may describe the interaction between Planck-sized pixels on the black hole event horizon. The effect, the curvature scalar is always zero, perhaps because there is no interaction between these pixels. From a historical point of view, the interpretation of the cosmological constant as pressure in 2009, the view of Jianyong Shen et al. no longer holds. The opinions of Mirza et al. and Ruppeiner have yet to be tested. There is no unanimous conclusion about the explanation of the constant zero of Ruppeiner's geometric curvature scalar.

{\bf Acknowledgements:}\\
This work is partially supported by  National Natural Science Foundation of China(No. 11873025).

\end{document}